\documentclass[aps,pra,twocolumn,superscriptaddress]{revtex4}
\usepackage{amssymb}
\usepackage[dvips]{graphicx}
\usepackage{graphicx}
\usepackage{amsmath}
\usepackage{subfigure}

\newcommand{\beq}{\begin{equation}}
\newcommand{\eeq}{\end{equation}}

\newcommand{\bea}{\begin{eqnarray}}
\newcommand{\eea}{\end{eqnarray}}

\begin{document}

\title{Ground state cooling is not possible given initial system-thermal bath factorization}
\author{Lian-Ao Wu}
\email{lianao_wu@ehu.es} \affiliation{Department of Theoretical
Physics and History of Science, The
Basque Country University (EHU/UPV),Post Office Box 644, 48080}
\affiliation{IKERBASQUE, Basque Foundation for Science, 48011, Bilbao, Spain}

\author{Dvira Segal}
\email{dsegal@chem.utoronto.ca}
\affiliation{Chemical Physics Theory Group, Department of Chemistry, University of
Toronto, 80 St. George St., Toronto, Ontario M5S 3H6, Canada }

\author{Paul Brumer}
\email{pbrumer@chem.utoronto.ca}
\affiliation{Chemical Physics Theory Group, Department of Chemistry, University of
Toronto, 80 St. George St., Toronto, Ontario M5S 3H6, Canada }

\maketitle

In the quantum regime, ground-state cooling of a small
object that is embedded in a thermal environment, such as
neutral atoms \cite{Dehmelt78}, ion traps \cite{Leibfried03},
mechanical resonators \cite{Schliesser08,Chan11}, nuclear spins
(polarization) \cite{Imamoglu}, is an intriguing challenge and one
of the most desirable of quantum technologies \cite{Review}. Mathematically, a
ground-state cooling (or polarization) process can be formulated as
a transformation from the initial state of the system+bath
to a final state, where the small object, the
``system", reaches its ground state. There have been variety of
ground state cooling schemes, for example, sideband cooling
\cite{Dehmelt78}, which have been carried out experimentally
\cite{Metzger04, Diedrich89}.
Here we prove that a fundamental constraint on the cooling dynamic implies that it is
impossible to cool, via a unitary system-bath quantum evolution,
a system that is embedded in a thermal environment down to
its ground state, if the initial state is a factorized product of system and bath states.
The latter is a crucial but artificial assumption often included in
many descriptions of system-bath dynamics  \cite{Breuerbook}.
The analogous conclusion holds for ``cooling" to any pure state of the
system.

To prove this fundamental statement we consider a generic arrangement with
a small entity, comprising a few degrees of freedom, referred to as the
``system" possibly subjected to time dependent fields,
interacting with a bath that is in a thermal equilibrium state.
The total Hamiltonian is given by
\bea
H=H_S+H_B+H_{SB},
\eea
where $H_S$ is the Hamiltonian of the system, $H_B$ is that of the
thermal bath, and $H_{SB}$ denotes the system-bath interaction
Hamiltonian. The details of these terms, whether controllable or uncontrollable (time-dependent or not),
do not alter our results. Time evolution in quantum mechanics is dictated by a propagator
$U(t_f,t_0)$, which transfers the full initial system+bath density matrix $\rho(t_0)$ to
the final density matrix $\rho(t_f)$,
\bea
\rho(t_f) =U(t_f,t_0)\rho(t_0)U^{\dagger} (t_f,t_0)
\label{eq:general}
\eea
Since $H$ is Hermitian, the unitary condition
$U(t_0,t_f)U^{\dagger}(t_f,t_0)=U^{\dagger}(t_f,t_0)U(t_f,t_0)=I$ is
satisfied, $I$ is the unity operator, and the trace of the density matrix $\rho$ is preserved.
Without loss of generality, the initial state of the total system is
assumed here to be diagonal,
$\rho(t_0)=\text{diag}(P_{0},P_{1},...)$.
If it is non-diagonal, we can diagonalize it by a unitary operator
$W$ such that $\text{diag}(P_{0},P_{1},...)=W\rho_{0}W^{\dagger}$.
The set $\left\{P\right\}$  corresponds to the eigenvalues
of $\rho(t_0)$, and we order the eigenvalues according to their
magnitude, $1>P_{0}\geqslant P_{1}\geqslant ...$.
Similarly, without loss of generality, we can also  assume a diagonal form for
the final state, $\rho(t_f) =\text{diag}(Q_{0},Q_{1},...)$, again
ordered as $1\geqslant Q_{0}\geqslant Q_{1}\geqslant ...$. The set
$\left\{Q\right\}$ corresponds to the eigenspectra of $\rho(t_f)$.
If $\rho(t_f)$ is not diagonal, it can be diagonalized with a
unitary matrix, $\text{diag}(Q_{0},Q_{1},...)=V^{\dagger }\rho(t_f)
V$. Overall, we can redefine the time evolution operator $U$ as
$VUW$, a unitary operator, to ensure that both initial state and
final state are diagonal matrices.
The density matrices $\rho(t_f)$ and $\rho(t_0)$, Hermitian
operators, are connected by a unitary (rotation) operation, thus
they must have identical eigenspectra, i.e., $\left\{
Q\right\} =\left\{ P\right\}$. Since the elements are ordered, we
can relate them one by one,
\begin{equation}
P_{k}=Q_{k}, \,\,\,\ k=0,1,2...
\label{eq:Q}
\end{equation}
We now define $d_0$ and $d_f$ as the number of non-zero eigenvalues
in the set $\{P\}$ and $\{Q\}$, respectively. Under a unitary
evolution, $d_0$ must be equal to $d_f$, a prerequisite for relation
(\ref{eq:Q}) to hold (or $d_{0}/d_{f}=1$ as $d_0$ goes to infinity).

Based on these simple considerations, we argue next that under system-bath unitary operations, acting on system or bath or both,
one {\it cannot} cool a mixed system state
down to its ground state if the total density matrix is	 initially
system-bath factorizable and the bath is thermal. That is, the process
\bea
\rho(t_0)=s\otimes b \stackrel{U} {\nrightarrow} |0\rangle
\langle 0| \otimes B=\rho(t_f) \label{eq:cool}
\eea
cannot be carried out with a unitary matrix $U$ (even if it operates on both the system and bath).
The left hand side in Eq. (\ref{eq:cool})
describes the initial system-bath product state. Here,
$s=\text{diag}(s_{0},s_{1},...)$ denotes the system density matrix
at $t_0$, which is anything but a pure state, and
$b=\text{diag}(b_{0},b_{1},...)$ denotes the bath state at that
time, a thermal state at nonzero temperature. The right hand side of Eq.
(\ref{eq:cool}) includes the target final state where the system has
been cooled down to its ground state $|0\rangle$ and the bath
is  a mixed state $B=\text{diag}(B_{0},B_{1},...)$ which is a diagonal
matrix that does not necessarily describe a thermal state. We now provide an
argument, which shows that one cannot evolve between these
initial and final states via unitary dynamics.

Define $N_S$ and $N_B$ as the Hilbert space dimension of the
system and bath, respectively. If the bath is initially thermal and
$s$ is not a pure state, the inequality $d_0>N_{B}$ holds.
In particular, if the system is initially thermalized we reach the
upper bound $d_0=N_SN_B$. On the other hand, the target state
$\rho(t_f)$ has only $d_f\leq N_{B}$ nonzero eigenvalues, reaching
the bound $d_f=N_B$ if the bath becomes  a thermal state at time
$t_f$. Since $d_f<d_0$, equation (\ref{eq:Q}), written here in the
form
\bea s_mb_{j}=B_k \label{eq:Q2} \eea
cannot be satisfied. Here the index $m$ counts the system eigenvalues,
$j$ and $k$ follow the bath eigenvalues. Hence, \emph{
system-bath unitary operations cannot cool a system connected to a thermal bath if the
system-bath initial state is factorizable and the system is initially in a
mixed state.}
The analogous proof holds for any final pure state of the system.

This proven conclusion stands in sharp
contrast to a multitude of studies, based on master equation
approaches, that demonstrate  ground state cooling from an initial
product system-bath state (e.g.,  Ref. \cite{Wilson-Rae04}).
While previous studies may have pointed out the unattainability of the absolute zero
in such situations \cite{Ketterle, Tannor,Mahler},
here we isolate the centrality of the factorization assumption,
and emphasize its strong implications regarding both the underlying physics
and the reliability
of master-equation type computational frameworks that often assume factorization.

It is of interest to examine a few related situations. First,
if the system is prepared in a pure state, we find that
$d_0=N_B$, and ground state cooling can potentially be performed
if Eq. (\ref{eq:Q2}) is satisfied.
Second, one can achieve ground state cooling by preparing the bath in a
non-thermal state. In this case we consider an initial
bath state $b$ with $N'_B$ nonzero eigenvalues, $N'_B<N_B$. This
results in $d_0>N'_B$ while $d_f\leq N_B$. 
These values could be made
identical if the states $b$ and $B$ are very different. As the
simplest example, consider both the system and the ``bath" as single
qubits, where initially the ``bath" populates its ground state,
$b_{0}=1$. Using Eq. (\ref{eq:Q2}), matching eigenvalues, we require
that $s_{0}=B_{0}$. System ground state can therefore be reached
here by the swapping operation.
A more involved scenario includes a two-qubit bath and a single-qubit
system where we initially set the system in a mixed state
while we prepare the bath in a non-thermal state with precisely
two zero eigenvalues, $b_2=0$ and $b_3=0$.
The prerequisite for ground state cooling, $d_0=d_f$,
could be fulfilled here
if at the end of the quantum evolution
all four bath eigenvalues $B_i's$ are made nonzero,
resulting in $d_0=d_f=4$.

Third, we note that  system-bath {\it correlated} initial
states \cite{discord} do allow for cooling. We illustrate this possibility  by
modeling the system as a qubit, with ground state $|0 \rangle$
and excited state $|1\rangle$. We construct the following
correlated initial state
\bea
\rho(t_0)=\left| 0\right\rangle \left\langle 0\right| \otimes b^{(0)}+\left|
1\right\rangle \left\langle 1\right| \otimes b^{(1)}
\eea
where $b^{(0)}=\text{diag} (b_{0},...,b_{n})$ and $
b^{(1)}=\text{diag} (b_{n+1},...,b_{N_{B}})$. As before, the target state is
$\rho(t_f)=|0\rangle \langle 0|\otimes B$. It is easy to confirm
that the prerequisite for cooling is satisfied, and the number of
non-zero eigenvalues for the initial and final density matrices is
identical, $d_0=d_f=N_B$. Furthermore, one could pair the
eigenvalues one by one, as required by Eq. (\ref{eq:Q}). For
example, we can set the system with $s_{0}=s_1=\frac{1}{2}$ and the
bath with $\frac{b_k}{2}=B_{k}$. As a result, the reduced density
matrix of the bath is the same, initially and finally, whereas the
reduced density matrix of the system at time $t_0$ is
$(\left|0\right\rangle \left\langle 0\right| +\left| 1\right\rangle
\left\langle 1\right| )/2$, for a given $n$ such that
$b_{0}+...+b_{n}=1/2$.
Note also that cooling may also be achieved when
measurement is involved in the cooling process \cite{Li11}.

Finally, consider {\it approximate} ground state cooling, defined as
the evolution from the initial state $\rho(t_0)=s\otimes b$  to the
final-factorizable state $\rho(t_f)=S\otimes B$,
where $S=\text{diag}(S_{0},S_{1},...)$, the diagonal
state of the system at the final time,
describes a system ``colder" than the initial one,
for example in the sense that $S$ has fewer nonzero elements than $s$.
This situation is typically assumed in the framework of Markovian master equations
\cite{Qbook}. 
Since the underlying quantum dynamics is unitary,
we should be able to match the
eigenvalues of the initial state and the final state.
In particular, the first two eigenvalues should fulfill $s_0b_0=S_0B_0$ and $s_0b_1=S_0B_1$.
The second relation holds as we assume
that the system energy gap, between its ground state and first
excited state, is larger than the corresponding gap in the bath,
$s_0b_1>s_1b_0$. These relations yield $B_{1}/B_{0}=b_{1}/b_{0}$,
translated to $e^{-(E_{1}-E_{0})/K_BT_{i}}=e^{-(E_{1}-E_{0})/K_BT_f}$ ($K_B$ is
Boltzmann constant),
if we further demand that the bath internal spectra is identical at
$t_0$ and $t_f$, and that the bath acquires a thermal equilibrium
state at the final time. Here $T_i$ and $T_f$ denote the temperature
at the different times. The last relation implies that the
final-time temperature is equal to the initial-time temperature,
i.e., the bath has not been changed through the cooling process, $\{b\}=\{B\}$.
As a result, to satisfy Eq. (\ref{eq:Q}), we must conclude that the system retains all
its values, $S_{m}=s_{m}$.  In the scenario described here,
quantum evolution {\it cannot} modify
the {\it system} population. Thus, even an approximate
cooling is impossible, as long as the system ground state is
nondegenerate.

In summary, ground state cooling  within system-bath unitary operations
is not possible given initial system-thermal bath factorization.
The linearity of unitary operations has, in the past, resulted in a no-go theorem,
the no-cloning theorem \cite{cloning}, one of the building blocks in modern
quantum information theory. Our no-go principle is similarly based
on unitary evolution, and it lays down the foundation for any theory
that aims at describing ground state cooling and pure state
preparation. For example, many recognized master equation techniques, as
well as Kraus operator based methods \cite{Breuerbook}, are predicated  on the
initial factorization of the system and bath. Adopting
these approaches to address issues of cooling should
be done with extreme caution, considering the fundamental constraint
exposed in this work.

\textbf{Acknowledgements}

\noindent L.-A. Wu has been supported by the Basque Government (Grant IT472-10), 
and the Spanish MICINN(Project No. FIS2009-12773-C02-02), DS acknowledges the 
NSERC discovery grant and PB's work has been supported by the U.S. Air Force Office of
Scientific Research under grant number FA9550-10-1-0260.



\end{document}